# Non-contact acoustic micro-tapping optical coherence elastography for evaluating biomechanical changes in the cornea following UV/riboflavin collagen cross linking: ex vivo human study

Mitchell A. Kirby[1], Ivan Pelivanov[1], Gabriel Regnault[1], John J. Pitre[1], Ryan T. Wallace[2], Matthew O'Donnell[1], Ruikang K. Wang[1,3], Tueng T. Shen[3,*]

[1]Department of Bioengineering, University of Washington, Seattle, Washington 98105, USA
[2]School of Medicine, University of Washington, Seattle, Washington 98195, USA
[3]Department of Ophthalmology, University of Washington, Seattle, Washington 98104, USA

[*]Correspondence: Tueng T. Shen, UW Medicine Eye Institute, Seattle, Washington 98104, USA.
Email: ttshen@uw.edu

**Abbreviations and Acronyms**

AµT (acoustic micro-tapping), OCT (optical coherence tomography), OCE (optical coherence elastography), NITI (nearly-incompressible transverse isotropy), IOP (intraocular pressure), CXL (cornea cross-linking), AAC (artificial anterior chamber), BSS (balanced saline solution)

**Financial Support**

This work was supported, in part, by NIH grants R01-EY026532, R01-EY028753, R01-EB016034, R01-CA170734, and R01-AR077560, Life Sciences Discovery Fund 3292512, the Coulter Translational Research Partnership Program, an unrestricted grant from the Research to Prevent Blindness, Inc., New York, New York, and the Department of Bioengineering at the University of Washington.

**Conflict of Interest**

No conflicting relationship exists for any author.

**Acknowledgements**

The authors wish to thank Kristine Sarmiento at the University of Washington for assistance in acquiring tissue samples. Kit Hendrickson is also acknowledged for assistance with figures.

**Data Availability**

The authors declare that all data from this study are available within the Article and its Supplementary Information. Raw data for the individual measurements are available upon reasonable request.

**Running head**

Non-contact OCE for measuring corneal elasticity in UV-CXL

**Conflict of Interest**

No conflicting relationship exists for any author.

**Acknowledgements**

The authors wish to thank Kristine Sarmiento at the University of Washington for assistance in acquiring tissue samples. Kit Hendrickson is also acknowledged for assistance with figures.

**Data Availability**

The authors declare that all data from this study are available within the Article and its Supplementary Information. Raw data for the individual measurements are available upon reasonable request.

**Running head**

Non-contact OCE for measuring corneal elasticity in UV-CXL

**Precis (35 words)**

Acoustic micro-Tapping Optical Coherence Elastography quantified changes in anisotropic elastic moduli of ex vivo human cornea treated with UV crosslinking. Two- and four-fold increases for in- and out-of-plane elastic moduli, respectively, were observed on average.

**Abstract (350 words)**

**Purpose.** To evaluate changes in the anisotropic elastic properties of ex vivo human cornea treated with UV cross-linking (CXL) using non-contact Acoustic micro-Tapping Optical Coherence Elastography (AµT-OCE).

**Design:** AµT-OCE was performed on normal and cross-linked human donor cornea in an ex vivo laboratory study.

**Subjects:** Normal human donor cornea (n= 22) divided into 4 subgroups. All samples were deemed unfit for transplant and stored in optisol.

**Methods.** Elastic properties (in-plane Young's, $E$, and out-of-plane, $G$, shear modulus) of normal and UV-CXL treated human corneas were quantified using non-contact AµT-OCE. A nearly incompressible transverse isotropic (NITI) model was used to reconstruct moduli from AµT-OCE data. Independently, cornea elastic moduli were also measured with destructive mechanical tests (tensile extensometry and shear rheometry).

**Main Outcome Measures:** Corneal elastic moduli (in-plane Young's modulus, $E$, in-plane, $\mu$, and out-of-plane, $G$, shear moduli) can be evaluated in both normal and CXL treated tissues, as well as monitored during the CXL procedure using non-contact AµT-OCE.

**Results.** CXL induced a significant increase in both in-plane and out-of-plane elastic moduli in human cornea. The statistical mean in the paired study (pre- and post-, n=7) of the in-plane Young's modulus, $E = 3\mu$, increased from 19 MPa to 43 MPa while the out-of-plane shear modulus $G$ increased from 188 kPa to 673 kPa. Mechanical tests in a separate subgroup support

CXL-induced cornea moduli changes and generally agree with non-contact AµT-OCE measurements.

**Conclusions.** The human cornea is a highly anisotropic material where in-plane mechanical properties are very different from those out-of-plane. Non-contact AµT-OCE can measure changes in the anisotropic elastic properties in human cornea as a result of UV-CXL.

The unique structure and organization of the collagen matrix within the cornea helps preserve shape and maintain transparency. Most corneal collagen is Type I, containing fibrils within the stroma that run the entire plane and connect with fibrils in the limbus or sclera.[1] Collagen fibrils are more tightly packed in the anterior portion compared to the posterior, but are generally arranged within the stroma in 200 - 300 stacked sheets (1-2 μm thick) referred to as lamellae.[2] The fibrils lie within a protein rich, hydrated proteoglycan mesh.[3,4] It is well established that the complex micro-network of collagen bundles contributes to macro-scale deformation and stability linked to functional vision. In pathologies affecting the cornea, microstructural changes in the collagen fibers and proteoglycan-rich fluid of the stroma alter the cornea's macro-scale biomechanical response to intraocular pressure[5] and can result in corneal shape changes and degraded visual acuity.[6–8]

In degenerative corneal diseases such as keratoconus, local changes in mechanical properties result in a bulging conical shape associated with vision loss.[7,9] Keratoconus is commonly treated with collagen cross-linking (CXL), a procedure reported to delay ectasia progression by increasing stiffness via ultraviolet (UV) light induced crosslinks in stromal collagen soaked in riboflavin. While cross-linking can halt disease progression by greatly reducing the number of required keratoplasties,[10] the mechanisms behind UV-CXL remain speculative[11] and both long-term and immediate outcomes remain unpredictable for individual patients.

Due to these uncertainties, a personalized biomechanical model predicting post-surgical corneal shape and function is needed to improve CXL outcomes. Personalized biomechanical models of the cornea can be generated if (i) cornea shape and thickness are mapped and (ii) corneal elastic moduli are known.

There are numerous clinical tools available to assess corneal shape, yet there are currently none that can measure corneal elastic moduli pre-, intra- and post-surgery in a non-destructive, non-invasive manner. While technologies based on tonometry (e.g. Ocular Response Analyzer (ORA), Dynamic Scheimpflug Analyzer (DSA)) have shown some success in correlating biomechanical estimates with disease progression,[12,13] these methods rely on a number of novel metrics that indirectly assess elastic moduli responsible for corneal deformation.[14–18] Specifically, both use a relatively high-pressure, widely distributed air-puff to deform the cornea and measure its response. While ORA has been used in multiple CXL studies in vivo, the one-dimensional nature of the system has resulted in poor sensitivity in detecting differences between CXL and control groups, where several studies described stabilized disease progression but found no change in the novel biomechanics-related metrics of corneal hysteresis (CH) or corneal resistance factor (CRF).[19,20] Additionally, tonometry-based techniques often require a non-trivial IOP correction in finite element simulations[14,21] and use a simplified isotropic mechanical model, leading to high variability depending on experimental conditions.

Recently, optical coherence elastography (OCE) has measured corneal stiffness at various length scales while simultaneously mapping shape in a single non-invasive measurement.[22] Contact-based OCE techniques have demonstrated potential to detect biomechanical differences between keratoconic and healthy corneas in vivo.[23] However, direct contact with the cornea is not ideal for clinical translation and cannot be performed during the cross-linking procedure.

Advances in both the sensitivity and imaging rate of phase-sensitive optical coherence tomography (PhS-OCT) have led to non-contact imaging systems that track propagating elastic waves launched by non-contact low-amplitude (less than a micrometer displacement) excitation.[24,25] Assuming an appropriate corneal mechanical model, dynamic OCE can measure in near real-time both the shape and biomechanical changes resulting from keratoconus and CXL surgery, with the potential for elastic moduli mapping at high spatial resolution.[26–29]

Until recently, the existing body of literature did not present a clear picture on the appropriate elastic model for human cornea. Common models assumed an incompressible, isotropic, and linear elastic material to describe corneal deformation, where a single parameter, the Young's modulus ($E$, or equivalently the shear modulus $\mu = E/3$), defined elasticity. However, the in-plane collagen arrangement in the cornea leads to an anisotropic biomechanical response,[30] i.e. corneal in-plane elastic properties are different from those out-of-plane, which suggests that the cornea more closely resembles a transverse isotropic material. As a result, up to three orders of magnitude differences have been reported from mechanical tests between the in-plane Young's modulus and the out-of-plane shear modulus depending on the method used.[30–33] Consequently, corneal elastic moduli obtained with isotropic models are highly inconsistent, greatly depend on the measurement technique and cannot be used to predict corneal deformation.[32,34–42] Because the anisotropic cornea cannot be characterized with a single elastic modulus, more appropriate models are required.

Recently, a model of a nearly incompressible transversely isotropic (NITI) medium was proposed based on corneal tissue microstructure.[30] It has been used to measure elastic anisotropy (i.e. in-plane, $\mu$, and out-of-plane, $G$, shear moduli) in porcine cornea using non-contact acoustic micro-tapping OCE (AµT-OCE). Notably, AµT-OCE is the only reported method that can quantify both $\mu$ (and therefore $E = 3\mu$) and $G$ simultaneously in a non-invasive, non-contact manner.[43] This is critical for in vivo studies of human corneal physiology and pathophysiology.

In this study, the NITI model was used to compute anisotropic mechanical moduli ($E$ and $G$) from elastic wave fields generated and detected using AµT-OCE in both ex vivo untreated human donor corneas and in samples subject to riboflavin/UV collagen crosslinking (UV-CXL) at physiologically relevant controlled pressures. Mechanical moduli were then directly measured in one subgroup of the same samples using destructive mechanical methods that loaded tissue in tension (in the plane of stromal fibers in tensile extensometry) and in shear (between the lamellar sheets in shear rheometry) to independently quantify $E$ and $G$. We show that the human cornea is highly anisotropic, and thus current isotropic models cannot predict corneal deformation under load.

Corneal mechanical anisotropy was also probed using AµT-OCE on a set of samples before undergoing UV-CXL and immediately after the procedure, as well as on a set of samples scanned during CXL, to both demonstrate potential in-line monitoring of corneal stiffness intraoperatively and understand the range of cornea moduli changes induced by CXL. The non-contact nature of AµT-OCE makes it promising to study corneal disease progression through objective measurements of corneal moduli. Indeed, it can play a critical role in optimizing CXL treatment of progressive keratoconus.

**Methods & Materials:**

**Cornea Preparation**

A total of 22 human corneal-scleral rings stored in Optisol (Chiron Ophthalmics, Irvine, California, USA) were secured through both the University of Utah Lion's Eye Bank (Murray, UT) and CorneaGen (Seattle, WA). Exclusion parameters included tissue over 2 months from enucleation, progressive corneal disease, and visual corneal damage. Such broad exclusion criteria were necessary to obtain research samples and resulted in a wide range of both ages and times between enucleation and analysis. Corneas were placed into three groups (Group A, B, and C) for different experiments based on their availability (**Fig. 1**). Institutional Review Board (IRB)/Ethics Committee approval was obtained.

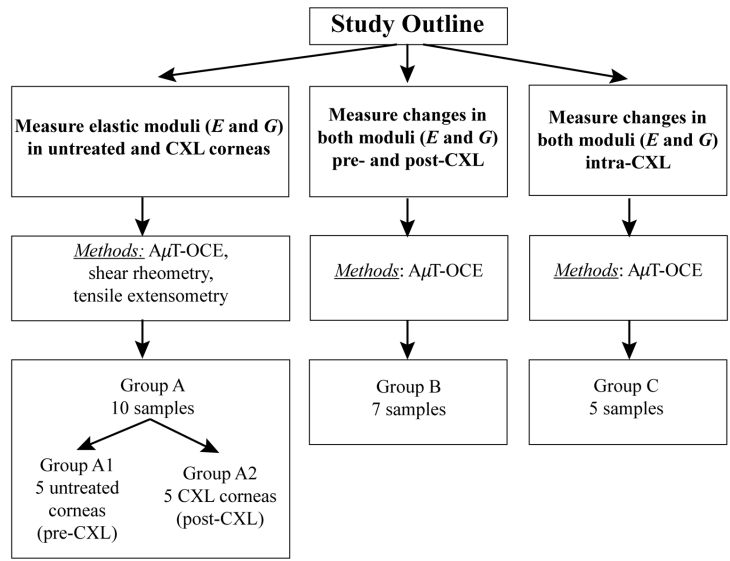

**Fig 1.** Schematic of the three different experiments performed and the corresponding number of samples in each group.

In Group A, 10 samples were subject to both non-contact AµT-OCE and destructive mechanical testing (shear rheometry and tensile extensometry) to determine mechanical moduli in both normal and cross-linked corneas. The results obtained were analyzed using different methods to justify the use of the NITI model (described below) in human cornea. All samples were removed from solution upon acquisition and fitted to an artificial anterior chamber (AAC) connected via a small tube to a bath filled with balanced saline solution (BSS). The bath could be raised and lowered relative to the cornea surface (according to a height previously calibrated using a digital hydrostatic pressure sensor) to control the intraocular pressure (IOP). Each sample was inflated to a desired pressure for 5 minutes, over which a single drop of BSS was applied to prevent corneal dehydration.

Corneas in group A were selected as untreated (group A1, n = 5) and cross-linked (group A2, n = 5) based on their availability. The average age of the untreated group (A1) was $53 \pm 5$ years and the samples were tested on average $31 \pm 18$ days from preservation. In the CXL group (A2), the average age was $29 \pm 10$ years and the samples were tested on average $15 \pm 5$ days from preservation.

In group A1, each cornea was scanned with AµT-OCE at physiologically relevant pressures (5, 10, 15, and 20 mmHg). Immediately following AµT-OCE scanning, each sample was transported in a BSS-damped cloth for parallel-plate rheometry at room temperature before sectioned into strips approximately 6 mm wide and transported for tensile testing.

In group A2, all samples were inflated to 15 mmHg and then subjected to a CXL procedure following the Dresden protocol prior to being scanned with AµT-OCE at pressures from 5 mmHg to 20 mmHg. Following CXL, corneas were transported for mechanical testing. Groups A1 and A2 independently demonstrated corneal anisotropy in both untreated cornea and those that underwent CXL.

In group B, n = 7 cornea were inflated to 15 mmHg and mechanical moduli were measured on the same sample using non-contact AµT-OCE before and after CXL. The average age was $51 \pm 17$ years and samples were scanned on average $30 \pm 5$ days from preservation. Group B provided a paired analysis of biomechanical changes on the same samples resulting from the CXL procedure.

In group C, n = 5 cornea were inflated to 15 mmHg and mechanical moduli were measured using AμT-OCE preoperatively, intraoperatively, and following CXL. The average age was 62 ± 8 years and samples were scanned on average 31 ± 15 days from preservation. Group C helped determine how quickly biomechanical changes occurred in each cornea during CXL.

**Collagen cross-linking (CXL) procedure**

The CXL procedure followed the widely adopted Dresden protocol.[44,45] A 0.1% riboflavin (RF) solution in 20% dextran was first prepared using 500,000 weight dextran (Alfa Aesar, Ward Hill, Massachusetts, USA) in isotonic saline. Before applying this solution, the epithelium was debrided using a blunt knife. Next, a 50 μL drop of riboflavin solution was placed on the cornea every 2 minutes over a total duration of 30 minutes. After riboflavin loading, each cornea was exposed to 3 mW/cm$^2$ of 370 nm (UV) light for 30 minutes. Riboflavin solution was reapplied every 5 minutes during the 30-minute irradiation period.

For the samples in group A2 (CXL group), a baseline AμT-OCE scan was taken prior to and post epithelial debridement. The pre- and post- epithelial debridement measurements showed no statistical difference. As such, the post-epithelial debridement scan was used as the 'pre-CXL' dataset. For samples in group B, (pre- and post- CXL), the pre-CXL scan was taken immediately following epithelium removal and prior to applying riboflavin/dextran. All corneas were then scanned following 30 minutes of riboflavin/dextran solution exposure and 30-minute UV light exposure.

Cornea samples in Group C were used to explore the rate of changes in cornea mechanical moduli during UV irradiation. In this experiment, a baseline scan was taken prior to and following epithelial debridement. As with Group A, the pre- and post- epithelial debridement measurements showed no statistical difference, and the post-epithelial debridement scan was considered the initial untreated dataset. Then, one 50μL drop of 0.1% RF in 20% dextran solution was placed on each cornea every 2 minutes for 30 minutes. Excess solution was removed, and a post-riboflavin scan was taken. Cornea were then exposed to 3 mW/cm² of 370 nm UV light for 30 minutes, and drops were continued every 5 minutes during irradiation. Excess solution was removed from the tissue surface and the cornea was scanned every 90 seconds during the 30-minute UV exposure. Following UV exposure, 5 post-CXL scans were performed.

**Nearly-incompressible transverse isotropic (NITI) model of corneal elasticity**

To account for cornea collagen microstructure consisting of thin layers (lamellas) primarily oriented in-plane of the stroma, the macroscopic stress-strain relationship in the cornea for small deformations is described by Hook's law in a NITI material, taking the form (in Voigt notation)[43]:

$$\begin{bmatrix} \sigma_{xx} \\ \sigma_{yy} \\ \sigma_{zz} \\ \tau_{yz} \\ \tau_{xz} \\ \tau_{xy} \end{bmatrix} = \begin{bmatrix} \lambda + 2\mu & \lambda & \lambda & & & \\ \lambda & \lambda + 2\mu & \lambda & & & \\ \lambda & \lambda & \lambda + \delta & & & \\ & & & G & & \\ & & & & G & \\ & & & & & \mu \end{bmatrix} \begin{bmatrix} \varepsilon_{xx} \\ \varepsilon_{yy} \\ \varepsilon_{zz} \\ \gamma_{yz} \\ \gamma_{xz} \\ \gamma_{xy} \end{bmatrix} \quad (1)$$

where $\sigma_{ij}$ denotes engineering stress, $\varepsilon_{ij}$ denotes engineering strain, $\tau_{ij}$ denotes shear stresses, $\gamma_{ij} = 2\varepsilon_{ij}$ denotes shear strains, and the subscripts $x$, $y$, and $z$ refer to standard Cartesian axes. Thus, there are four independent elastic constants, $\lambda$, $\delta$, $G$, and $\mu$, defining the cornea's elasticity matrix. Because the cornea is nearly-incompressible, the longitudinal modulus, $\lambda$, does not influence deformation. $G$ and $\mu$ represent out-of-plane and in-plane shear moduli, respectively, whose magnitudes were shown to differ in corneal tissue. Indeed, several studies (including OCE,[30,43] and mechanical testing[32,46,47]) have reported the out-of-plane shear modulus, $G$, to be on the order of tens of kPa, whereas the in-plane shear modulus, $\mu$, to be on the order of MPa.[40,43]

The in-plane Young's modulus

$$E_T = 3\mu + \frac{\mu\delta}{4\mu+\delta} \qquad (2)$$

is defined by $\mu$ and $\delta$, but does not depend on $G$. In Refs.[43,48] we showed, that

$$-2\mu < \delta < 0 \qquad (3)$$

to obey cornea symmetry but highlight that $\delta$ cannot be extracted from guided wave propagation. Fortunately, constraints on $\delta$ limit the range of the in-plane Young's modulus to

$$2\mu < E_T < 3\mu. \qquad (4)$$

Here we assume that

$$E = E_T \cong 3\mu, \qquad (5)$$

producing slight overestimates of actual values. Note that due to strong corneal anisotropy, $\delta$ likely varies little between corneas in a population. The exact relationship between $\mu$ and $E$ can be clarified with tensile testing and is the focus of future studies. Even with this approximation (Eq. 5), the elastic moduli $E$ and $G$ can fully characterize corneal deformations for small deformations.

**Acoustic Micro-tapping Optical Coherence Elastography (AµT-OCE)**

Elastic moduli of human cornea were quantified using wave fields of guided mechanical waves generated and measured with a phase-sensitive spectral-domain OCT system combined with an air-coupled AµT ultrasound transducer. The OCT system used polarization-maintaining optical fibers and components, as described in detail previously.[43]

To generate and track propagating elastic waves, the OCT system operated in M-B-mode, in which a sequence of 512 A-scans were repeated at the same location, (referred to as an M-scan). A trigger signal was sent to a function generator at the beginning of each M-Scan to direct a 100 µs-long chirped (1 MHz-1.1 MHz) waveform to an air-coupled piezoelectric transducer (AµT source) with a matching layer, providing a temporally localized and spatially focused acoustic 'push'.[26] The acoustic force created a localized displacement (hundreds of nanometers in amplitude) at the cornea's surface. The pulse/M-scan sequence was captured at 256 spatial locations across the imaging plane to create a 3-D volume (z, x, t). A complete M-B–scan consisted of 1024 depth × 256 lateral locations × 512 temporal frames (captured at a 46 kHz frame rate) with an effective imaging range of 1.5 mm × 10 mm x 11 ms (axial × lateral × temporal). Each scan took ~3 seconds to acquire and save. Note that the OCT scanning range of 10 mm in the lateral direction was partially shaded (~1.5 mm from the first scan location) by the AµT-OCE transducer.

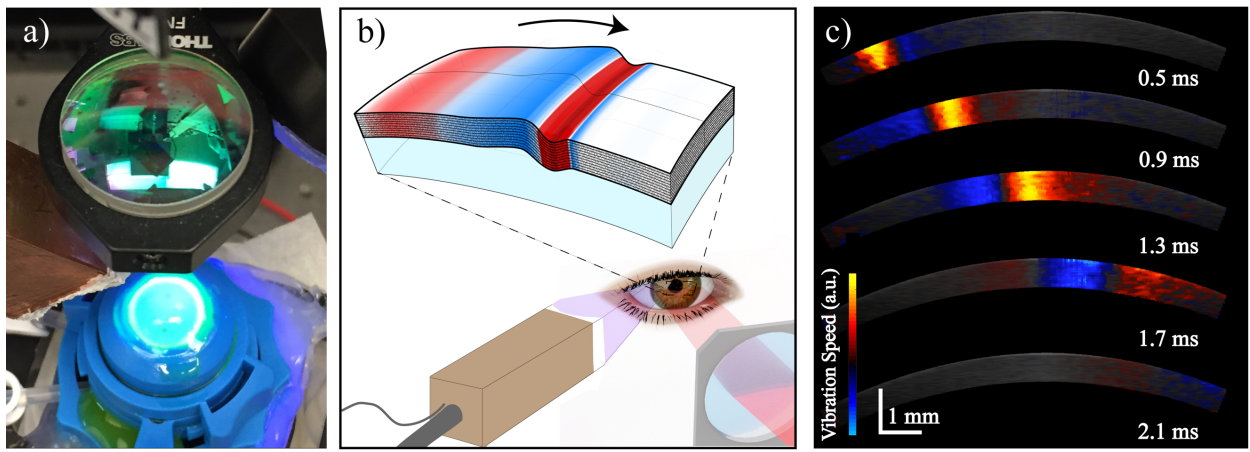

**Fig. 2.** a) AµT-OCE imaging system with cornea inflated via artificial anterior chamber undergoing UV- CXL treatment. b) Schematic of mechanical wave propagation within the cornea. c) Snapshots of elastic transients detected at different time instants using phase-sensitive OCT.

The resulting three-dimensional dataset was used to reconstruct central corneal structure pre- and post- CXL, as well as track propagating mechanical waves in the cornea following AµT excitation (**Fig. 2**). The central corneal thickness pre- and post- CXL was calculated from the OCT structural image after correcting for the cornea's refractive index (assumed to be 1.38) (**Figs. 3a,d**). The OCT-measured local particle axial vibration velocity ($v_z(x,z,t)$) was obtained from the optical phase difference $\Delta\varphi_{opt}(x,z,t)$ between two consecutive A-line scans at each location.[25] Surface vibrations were detected using automatic segmentation of the anterior surface based on an edge detection algorithm. One half of a Gaussian window (HWHM = 90 µm, weight decreasing with depth) was applied as a weighted-average to phase data in the anterior 183 µm of the cornea, resulting in robust sampling of the vertical displacement for propagating guided elastic waves along the air-tissue boundary. Guided wave signals in the cornea recorded at different time moments (see **Fig. 2c**) were combined to form a wave field (**Figs. 3b,e**).

A temporal super Gaussian filter ($SG$) that followed the maximum vibration velocity of the wave-field $t_m^{wf}(x)$ at each discrete position $x$, was then applied:

$$SG(t) = \exp\left[-\left(\frac{1}{2}\left(\frac{t - t_m^{wf}(x)}{\sigma_t}\right)^2\right)^2\right] \qquad (6)$$

with $\sigma_t$ = 0.5 ms, and a two-dimensional Fourier transform was computed to describe wavefields in the frequency-wavenumber (*f-k*) domain (**Figs 3c,f**).

A solution to the equation for guided wave propagation in the NITI material[30] was found that most closely matched the experimental *f-k* spectrum using an iterative routine that converged on a best-fit of the dispersion relation. By varying both the in-plane ($\mu$, assuming tensile isotropy where $\mu = E/3$) and out-of-plane, $G$, shear moduli and using the corneal thickness, $h$, as a constraint, a dispersion relation that closely matched experimental data was determined (see **Supplemental Methods 1, 2** for details on the fitting routine). In all samples, the wave energy was mainly contained within the $A_0$ mode[22,30] and fit to a theoretical solution of the $A_0$ mode assuming a linear-elastic NITI material.

Due to the nature of the weighted fitting routine, residuals are not generated, making traditional confidence interval methods difficult to apply. As such, asymmetrical error bars corresponding to uncertainties in $E$ and $G$ for each cornea (due to both model and data error) are described in **Supplemental Methods 2**. All cases that produced a poor fit quality (detailed in **Supplemental Methods 3**) were omitted.

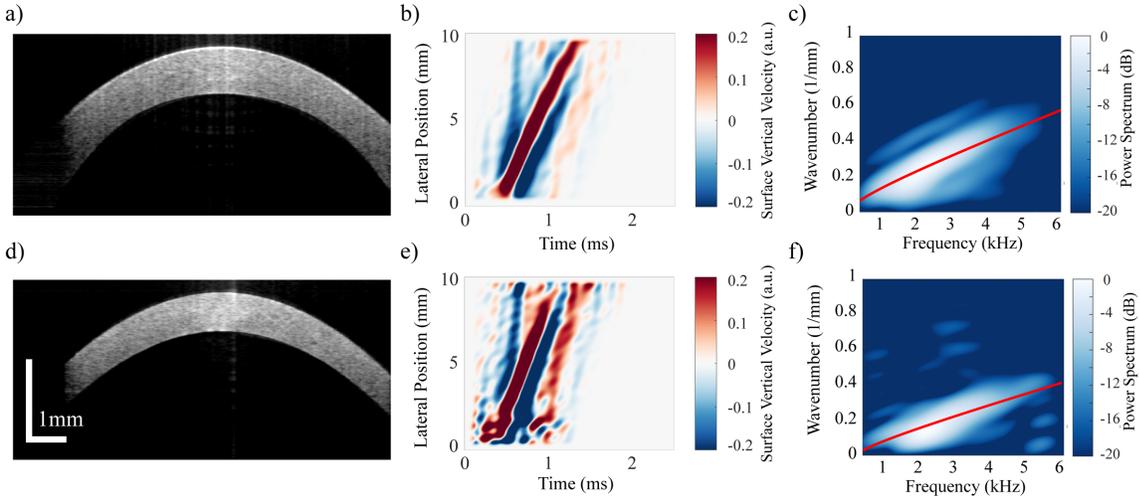

**Fig. 3.** Central cornea cross-section and automatic segmentation to determine the surface location and thickness a) pre- and d) post-CXL. Wave fields of propagating guided waves in the same cornea b) pre- and e) post-CXL tracked over ~8.5 mm of corneal tissue. Best-fit solution to the dispersion equation (based on a unique combination of elastic moduli, $E$ and $G$, displayed in red) on top of the measured waveform in the 2-D Fourier spectrum for the corresponding cornea c) pre- and f) post-CXL.

**Destructive biomechanical testing**

Following AµT-OCE, sub-groups A1 and A2 were subject to mechanical testing under tensile and shear loading.

The frequency-dependent shear behavior of a subset of corneal buttons was measured using a rheometer (Anton Paar MCR 301 Physica, **Fig. 4a**) over a range of 0.16-16 Hz using a 5 N compressive preload and a peak shear strain of ~.1%. Rheometry produced estimates of the frequency-dependent storage ($G'(\omega)$) and loss ($G''(\omega)$) components, corresponding to the real and imaginary parts of the modulus $G$.

Each sample was then cut into strips for tensile testing and pneumatically clamped (2752-005 BioPuls submersible pneumatic grips, 250 N max load) to the edge of the cornea (**Fig. 4b**). A 50 mN pre-load was applied and each sample was stretched at 2 mm/min (Instron model 5543) up to either 10% strain, 10 N load, or corneal breakage (whichever occurred first). Two load-unload cycles were performed to precondition the tissue, followed by 3 rounds of force-elongation followed by relaxation. The force-elongation measurement was converted to stress-strain according to sample geometry. A second order exponential was fit using three-sets of raw data to determine the stress-strain curve. The in-plane Young's modulus, $E$, was defined as the tangential slope of the stress-strain curve. Consequently, extension testing provided a value for the strain-dependent Young's modulus, $E$, up to 10% strain.

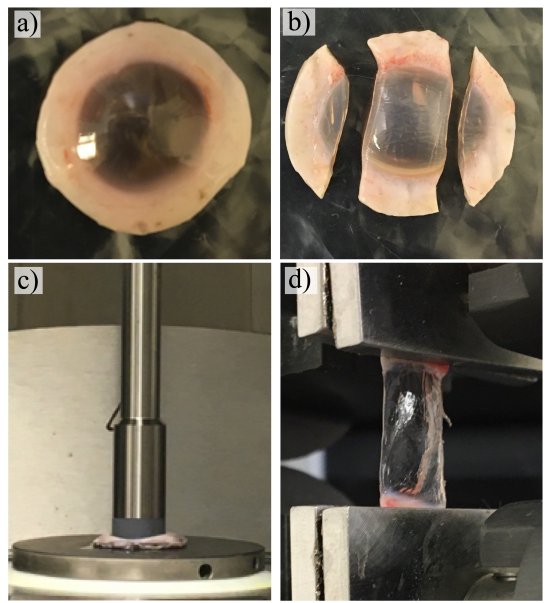

**Fig. 4.** a) Cornea buttons prepared and c) loaded in parallel-plate rheometry to measure out-of-plane shear modulus, $G$. b) Cornea strips prepared and d) clamped during extension testing to determine in-plane Young's modulus, $E$.

## Results

### Elasticity changes with IOP in untreated and CXL cornea

*Corneal Thickness*

The mean thickness of all samples in Group A is presented in **Table 1** for all tests. The CXL group (A2) was statistically thinner ($p<.05$) than the untreated group (A1) at all pressures tested with OCE and during rheometry. Corneal thickness in AµT-OCE was determined by OCT imaging; in rheometry - via parallel plate separation; and in extension testing - using a digital micrometer. Note that the OCE set-up placed the cornea under inflating forces and assumed a refractive index, rheometry required a flattening of the cornea and was insensitive to differences in the central and peripheral thickness, and the digital micrometer required hand-tuning.

**Table 1.** Corneal thickness for each testing condition.

| | Mean thickness [µm] (± standard deviation) | | | | | |
|---|---|---|---|---|---|---|
| | OCE | | | | Rheometry | Extension Testing |
| | 5mmHg | 10mmHg | 15mmHg | 20mmHg | | |
| **Group A1 (Untreated)** | 839 (±230) | 821 (±219) | 825 (±260) | 816 (±259) | 852 (±94) | 781 (±161) |
| **Group A2 (CXL)** | 630 (±114) | 633 (±120) | 631 (±123) | 633 (±128) | 654 (±107) | 645 (±129) |

Because OCE enabled measurements in cornea inflated to different IOP, a general (but not statistically significant) decrease in thickness was measured as pressure increased in the untreated group. Inflation induced thickness changes were not corrected for in rheometry and extension testing due to limitations in the experimental set up.

*Out-of-plane shear modulus, G*

In the 'untreated' group (A1), the OCE-determined out-of-plane shear modulus, $G$, for each sample is shown in **Fig. 5a**. The statistical mean and standard deviation of the moduli between samples was 70 kPa ($\pm$ 29 kPa) at 5 mmHg, 105 kPa ($\pm$ 55 kPa) at 10mmHg, 169 kPa ($\pm$ 129 kPa) at 15 mmHg, and 240 kPa ($\pm$ 191 kPa) at 20 mmHg. In the 'CXL' group A2, the OCE-measured out-of-plane shear moduli, $G$, are also shown in **Fig. 5a**, where the statistical mean and standard deviation of the moduli were 151 kPa ($\pm$ 67 kPa) at 5 mmHg, 242 kPa ($\pm$ 76 kPa) at 10 mmHg, 290 kPa ($\pm$ 78 kPa) at 15 mmHg, and 334 kPa ($\pm$ 99 kPa) at 20 mmHg.

A relative change of modulus, $G$, with increasing IOP for both untreated and CXL groups can be seen, where the mean shear modulus, $G$, increased at 20 mmHg (relative to the value at 5 mmHg) by a factor of 3.4 in the normal group compared with 2.2 in the CXL group.

Mechanical measurements using parallel-plate rheometry are shown in **Fig. 5b** for both untreated and cross-linked corneas. In untreated samples, the statistical mean and standard deviation of the moduli between samples were 201 kPa ($\pm$ 55 kPa) at 0.16 Hz and 336 kPa $\pm$ (56 kPa) at 16 Hz. For the cross-linked group, the statistical mean and standard deviation were 270 kPa ($\pm$ 34 kPa) at 0.16 Hz and 388 kPa ($\pm$ 54 kPa) at 16 Hz. Rheometry measurements lie within the range of OCE-determined modulus, $G$.

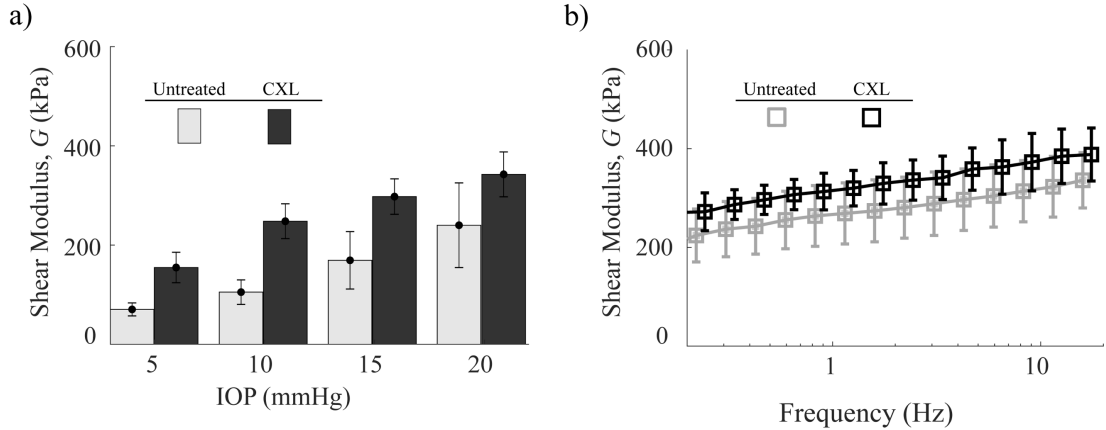

**Fig. 5.** a) Out-of-plane shear modulus, $G$, in untreated (A1) and CXL (A2) cornea groups measured with AμT-OCE at intraocular pressures from 5-20 mmHg respectively. b) Shear storage modulus, $G$, measured with parallel plate rheometry for untreated (A1) and CXL (A2) cornea groups.

Individual results from each cornea have been included in **Supplemental Methods 4**. Note that OCE measured values in individual corneas generally tracked with rheometry measurements. Uncertainty intervals and detailed exclusion criterion can also be found in the **Supplemental Methods**.

OCE-measured values of $G$ were statistically different (p<0.05) between untreated and CXL groups at 5 mmHg and 10 mmHg, but not at 15 mmHg (p=.11) and 20 mmHg (p=.36). The means of rheometery results were not statistically different between groups at any frequency.

*In-plane Young's modulus, E*

AμT-OCE experiments simultaneously evaluated in-plane, $\mu$, and out-of-plane, $G$, shear moduli. The modulus $\mu$ can be then converted to the in-plane Young's modulus ($E = 3\mu$).[28,45] Note that $E$ cannot be determined from rheometry measurements. We used tensile extensometry to compare $E$ with that determined from AµT-OCE.

In the untreated group (A1), the OCE-measured in-plane Young's modulus, $E$, for each sample is shown in **Fig. 6a**. The statistical mean and standard deviation of the moduli between samples were 15 MPa ($\pm$ 5 MPa) at 5mmHg, 21 MPa ($\pm$ 8 MPa) at 10 mmHg, 24 MPa ($\pm$ 9 MPa) at 15 mmHg, and 24 MPa ($\pm$ 13 MPa) at 20 mmHg. In the CXL group (A2), each of the in-plane Young's moduli, $E$, is also shown in **Fig. 6a**, where the statistical mean and standard deviation of the moduli between samples were 37 MPa ($\pm$ 7 MPa) at 5 mmHg, 39 MPa ($\pm$ 7 MPa) at 10 mmHg, 44 MPa ($\pm$ 6 MPa) at 15 mmHg, and 47 MPa ($\pm$ 7 MPa) at 20 mmHg.

Relative changes in Young's modulus with IOP were smaller for CXL corneas compared to that for untreated ones where Young's modulus for untreated corneas increased by a factor of 1.6 with an increase in IOP from 5 mmHg to 20 mmHg and $E$ increased by ~1.3 for CXL cornea.

The in-plane Young's modulus determined by extension testing at 1-10% strains following dissection of the samples is shown in **Fig. 6b**, where the statistical mean and standard deviation at 1% strain were 3.2 MPa ($\pm$ 2.2 MPa) and at 10% strain were 51 MPa ($\pm$ 31 MPa). In the CXL group (A2), the in-plane Young's modulus determined by extension testing at 1-10% strains is also shown in **Fig. 6b** where the statistical mean at 1% strain was 2.9 MPa ($\pm$ 0.9 MPa). The high strain modulus was either at 10% strain, or at the inflection point before structural damage occurred. The high strain modulus in the CXL group was 74 MPa ($\pm$ 50 MPa). Samples in the CXL group demonstrated significant tissue damage at higher strains and required more force to extend.

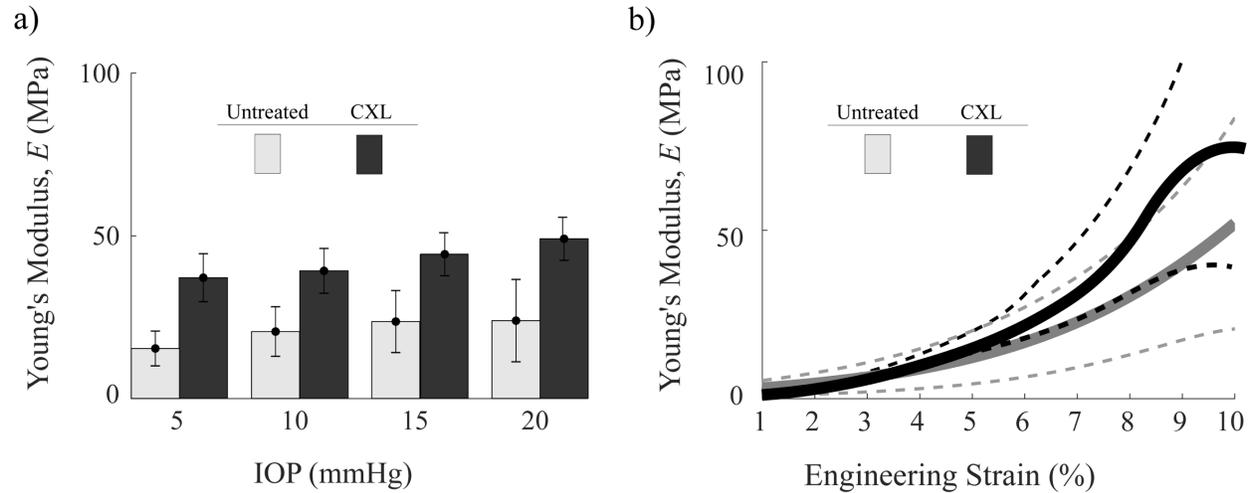

**Fig. 6.** a) In-plane Young's modulus, $E$, in untreated group (A1) and CXL group (A2) measured with AµT-OCE at intraocular pressures from 5-20 mmHg. b) Strain-dependent Young's moduli measured via extension testing up to 10% strain in untreated group (A1) and CXL group (A2). $E$ was determined via extension testing to 10% strain, or where visible tissue damage occurred. The dashed lines correspond with plus or minus one standard deviation. Individual tests can be found in **Supplemental Methods 4**.

OCE-measured values of $E$ were statistically different between groups at all pressures ($p<0.1$ for IOP of 5 mmHg to 20 mmHg, respectively.) The means of the best fit curve of the strain dependent tangential modulus were not statistically different across all strains. Uncertainty intervals and individual results from each cornea have also been included in **Supplemental Methods 4**.

The independent sub-group study suggests that in general, (i) both moduli ($G$ and $E$) could be measured in ex vivo human cornea using the NITI model in untreated and CXL treated samples (ii) OCE measurements generally agree with mechanical tests. Thus, the effect of CXL on corneal elasticity can be explored in more detail without direct comparison to mechanical tests.

This result also suggests that OCE can monitor elasticity changes throughout the entire treatment cycle, including untreated corneas pre-, intra- and post-CXL.

Additionally, the results suggest that one-to-one comparisons before and after CXL are required to better understand the effect of UV-CXL. Such scans can measure CXL-induced changes for each individual cornea rather than relying on group comparisons between untreated and treated samples. Analyzing results for every cornea independently should greatly reduce the effect of population variability of corneal properties, which is key to developing personalized models of the eye for optimal surgery planning. Results on ex vivo preliminary studies are considered in the sub-sections below.

**Independent monitoring of corneal moduli pre- and post- CXL**

Results for each sample in Group B, where moduli were quantified immediately before and after cross-linking, can be seen in **Fig. 7a** and **Fig. 7b.** Clearly, changes induced by CXL vary from cornea to cornea, suggesting that individual treatment plans may be required to reach desired moduli outcomes.

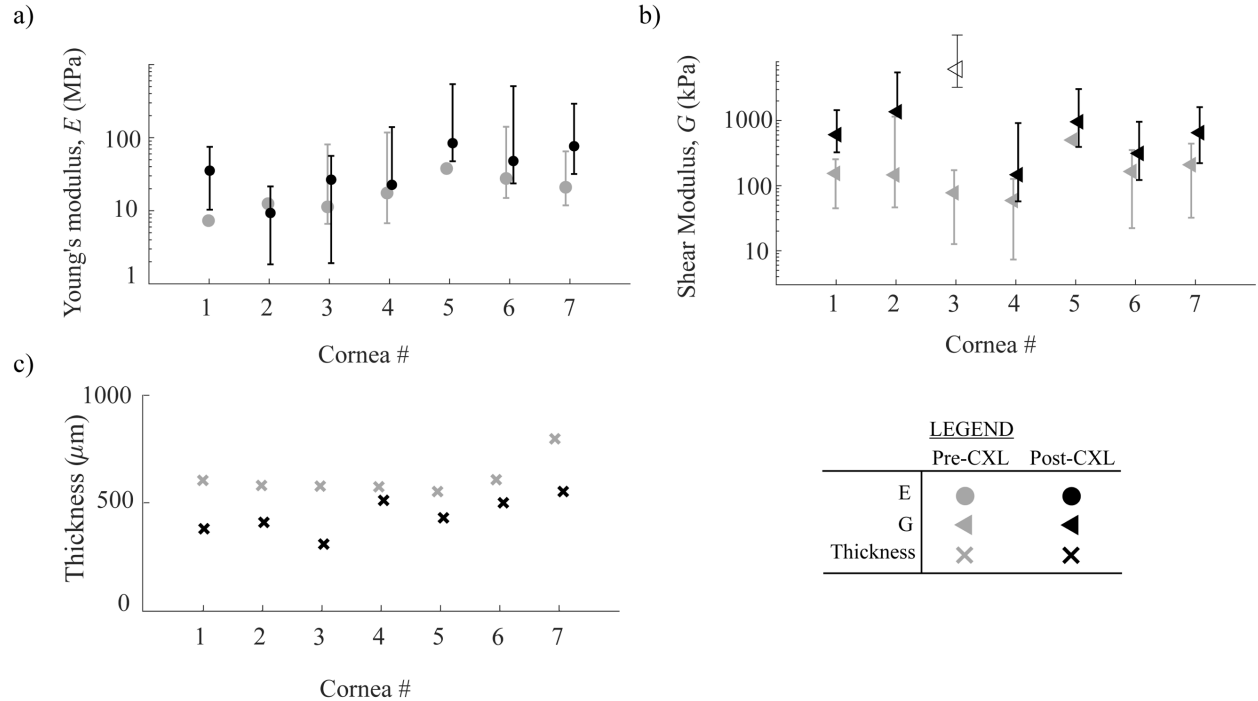

**Fig. 7.** CXL-induced differences in a) in-plane Young's modulus, $E$, and b) out-of-plane shear modulus, $G$, for cornea in group B. Error bars correspond with uncertainty intervals described in **Supplemental Methods 2**. Note that in one sample (cornea #3) all 5 scans demonstrated high uncertainty in $G$ following CXL, thus the value is displayed as an 'open' triangle and omitted from analysis of the means. c) Thickness for each individual cornea pre- CXL and post- CXL.

The statistical mean (standard deviation) of $E$ changed from 19 MPa ($\pm$ 4 MPa) to 43 MPa ($\pm$11 MPa) and the shear modulus, $G$, from 188 kPa ($\pm$ 148 kPa) to 673 kPa ($\pm$ 440 kPa) (see **Fig. 8).** Note that in cornea #3, (post-CXL), (**Figure 7b**), all 5 repeat scans demonstrated high uncertainty in $G$. As such, its value was omitted from the calculation of mean values.

While both moduli increased following UV exposure in RF/dextran-soaked cornea, the average change in in-plane Young's modulus, $E$, was approximately double (2.3), whereas the out-of-plane shear modulus $G$ changed by more than three times (3.6). Notably, both $E$ and $G$ increased for all samples following CXL.

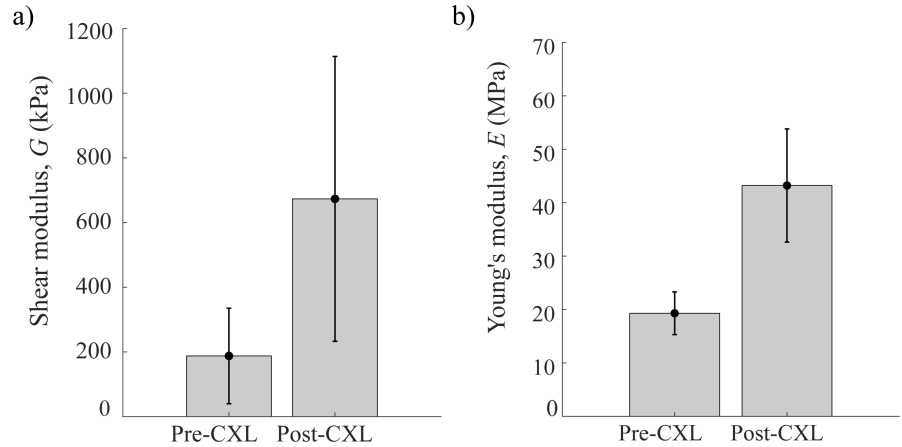

**Fig 8.** The mean and standard deviation in a) $G$ and b) $E$ prior to and following UV-CXL in n=5 samples.

Both $E$ and $G$ had significantly different stiffening responses to CXL, where statistical significance (p < 0.05) was determined using a two-tailed t-test where the null hypothesis was no difference between the means in the pre- and post -CXL values. The p-value for $G$ was 0.03 and for $E$ was 0.02, indicating a significant stiffening response in both moduli.

The central corneal thickness following epithelium removal was calculated from the OCT structural image pre- and post- CXL. The measured thickness pre-CXL was 577 μm ($\pm$ 58 μm) and reduced to 450 μm ($\pm$ 80 μm) post-CXL. The measured thickness was also statistically different in the samples following CXL (p = 0.001).

**Evaluation of both corneal moduli during CXL procedure**

To demonstrate the ability of AµT-OCE to directly quantify corneal biomechanics intraoperatively, $E$, $G$, and thickness were measured in group C every 90 seconds during the procedure. A moving average on both the output value and uncertainty intervals was performed over 4.5 minutes (the full data-set can be found in **Fig. S10**). The mean and standard deviation of the relative modulus (i.e., modulus at any time normalized to the modulus pre-CXL) can be seen in **Fig. 9**, along with a linear fit (over time) to highlight individual changes in $G$, $E$, and central cornea thickness with CXL. While both moduli increased with UV exposure, $E$ changed on average by about 2 times (from 15 MPa ($\pm$ 4 MPa) to 31 MPa ($\pm$ 6 MPa)) whereas $G$ changed by approximately 4 times (from 107 kPa ($\pm$ 26 kPa) to 421 kPa ($\pm$ 110 kPa)) following CXL. Both moduli changed significantly ($p<.01$) between pre- and post-CXL values. Each cornea changed stiffness and thickness at a slightly different rate. Non-negligible stiffening also occurred from the RF/dextran alone within the first 30 minutes. Note that two (2) of the corneas resulted in a poor fit for $G$ over the last 6 minutes. Four (4) of the 5 samples provided sufficient fit for $G$ in the scans taken immediately following CXL.

The cornea progressively thinned during crosslinking. The measured thickness pre- and post- CXL was 560 µm $\pm$ 70 µm and 340 µm $\pm$ 20µm, respectively. Combined elasticity and thickness changes during CXL will have significant impact on both the overall corneal stiffness and potential refractive changes. Although this experiment required a slight modification of the FDA-approved Dresden protocol (i.e. removal of RF/dextran from the corneal surface prior to each scan), the results indicate that AµT-OCE potentially can monitor treatment progression.

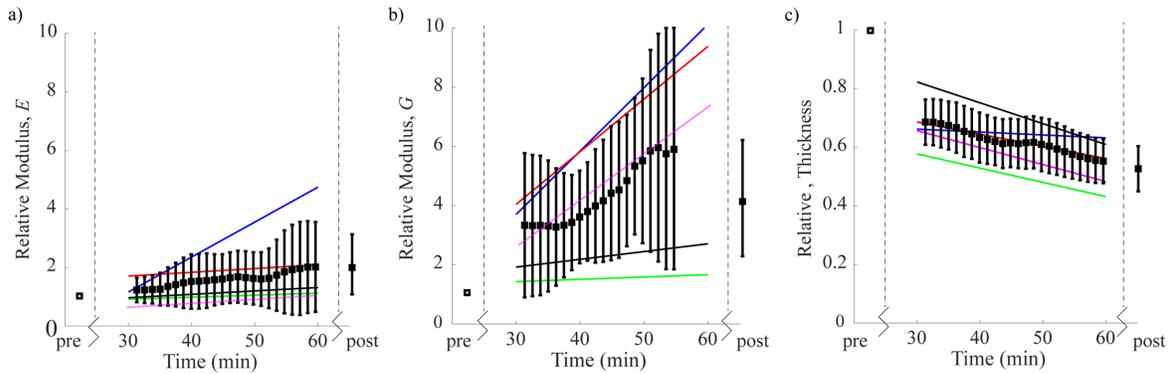

**Fig. 9.** a) Young's Modulus E and b) Shear Modulus G, relative to the pre-CXL value; c) central cornea thickness relative to the pre-CXL value. Black markers are the mean and standard deviation of the relative changes between the n=5 samples. The solid colored lines represent a linear best-fit of the relative change from each individual cornea. For 2 of the corneas, the fit for $G$ was very poor over the last 6 minutes, so the values have been omitted from the displayed average. The post CXL average includes 4 corneas.

**Discussion:**

Mechanical anisotropy was confirmed in ex vivo human cornea, where the in-plane tensile (Young's) modulus, $E$, was on average multiple orders of magnitude larger than the out-of-plane shear modulus, $G$. Similar to porcine cornea,[43] the results of Group A suggest that human cornea can be approximated as a nearly incompressible transversely isotropic (NITI) material. Using the NITI model to describe corneal deformation plays a key role in explaining the order(s) of magnitude difference in corneal stiffness estimates extracted from shear- and tensile-based mechanical measurements.

Guided elastic wave propagation mixes fundamental modes allowing $E$ and $G$ to be measured simultaneously in the cornea.[30] As shown previously for strongly anisotropic materials, a change in $E/\mu$ results in a small, but detectable change in $A_0$-mode dispersion, particularly in the low frequency region (where the wavelength is long compared to the thickness). As such, the wave velocity is roughly dominated by the shear modulus, $G$, and applying an isotropic model to fit the same recorded waveforms of guided mechanical waves would lead to a rough approximation of $G$, not $E$ or $\mu$. In this case, Young's modulus, $E$, calculated using the expression $E = 3G$ (as done previously[49–51]) would be highly underestimated compared to the correct value of $E = 3\mu$. Such an underestimated $E$ modulus used in prediction models would produce an inaccurate estimate of cornea stiffness, its deformation, and would fail to evaluate clinical procedure outcomes, including that for CXL.

To better understand the limitations in quantifying moduli, a method to determine uncertainty intervals in OCE-measured modulus was presented in **Supplemental Methods 2** using a goodness-of-fit routine that compared the $A_0$ structure for different values of $E$, and $G$, independently. As demonstrated previously, the $A_0$ mode is increasingly insensitive to changes in $E$ as the degree of mechanical anisotropy increases ($E >> G$).[30] This suggests that when $E$ is much larger than $G$, it can be very difficult to accurately determine $E$ for typical SNR's achieved with OCE. Signal averaging or multi-channel scanning may be used to improve SNR and enable more accurate determination of $E$, but can be difficult to perform in vivo. To predict refractive changes and model the cornea as a curved lens system, both $G$ and $E$ are required, yet, one question that remains is the precision needed to measure a change in $E$ corresponding to a physical change in corneal shape. Static models are required to answer this question and remain an area of future interest.

The results of Group A suggest that AµT-OCE can quantify both $G$ and $E$, with less certainty around $E$ (**Supplemental Method 2**). Although $E$ in the NITI model can have a high degree of uncertainty, analyzing guided wave propagation remains the only non-contact, non-destructive methodreported to date to estimate in-plane tensile stiffness.

While elastic moduli measured via OCE and mechanical methods in this study generally agree, there are subtle experimental differences that make it difficult to quantitatively compare values between methods. For example, an artificial anterior chamber (AAC) was used to inflate corneal samples, providing additional boundary conditions in the lateral direction limiting the propagation distance of the guided wave over the cornea, that differ from those of both in vivo and destructive ex-vivo settings. While probing tissue inflated via an AAC is more like cornea loading in vivo, mechanical boundary conditions are markedly different from those for whole-globe samples. Together with AµT excitation, it can lead to modifications in the low-frequency region of the *f-k* spectrum, worsening fitting accuracy. Additionally, to facilitate extension testing, a portion of the sclera was left on each sample during rheometry, which was shown to induce a measurable increase in the apparent modulus $G$.[43] Sample transportation, time limitations, and measurement precision also contributed to differences in measured thickness between OCE and destructive methods, potentially influencing final quantitative moduli values.

In Group B, both the out-of-plane shear modulus, $G$, and in-plane Young's modulus, $E$, increased significantly following cross-link formation. The measured increase in both $G$ and $E$ suggests that CXL strengthened the tensile and flexural response, as well as the ability of the cornea to resist shearing forces. The measured increases in $G$ and $E$ of approximately 4-fold and 2-fold, respectively, generally agreed with values previously reported. For example, Marbini et. al.[52] and Del Buey et. al.[53] reported a post-CXL increase in corneal Young's modulus, $E$, by a factor of ~1.5-2 in the 7-10% strain range, while Wollensak et al. report an increase of a factor of 4.5 (1.3 MPa- 5.9 MPa at 6% strain) in human cornea tested via strip extension within 1 hour of enucleation.[36] The results of this study are also consistent with Aslanides et. al.,[41] who reported an increase as high as 20-fold in the shear modulus $G$ in porcine cornea following CXL (as measured using a parallel-plate scheme) and Sondergaard et. al.[42] who reported a general 2-5 fold increase in $G$ in human cornea under 5% compressive strain, where samples were probed using a parallel plate arrangement to provide a translational shear (as opposed to rotational).

Following UV-CXL, the out-of-plane shear modulus, $G$, and the in-plane Young's modulus, $E$, increased from untreated values by different multiples. Specifically, the shear modulus, $G$, (most likely related to the proteoglycan mesh and interlamellar connectivity) changed almost four times in CXL corneas compared to untreated corneas, whereas the tensile modulus, $E$, changed only about two times. This result suggests that CXL changes elastic properties of the connective tissue mesh between the lamellar sheets in addition to that of fibers, making the connective tissue or proteoglycan mesh apparently 'stickier'. The effect of the connective matrix on mechanical stability is well known in different material science and engineering areas, where fibers embedded in a composite material can produce almost equivalent tensile properties as the fibers, with shear properties defined independently by the connective matrix. Consistent with this effect, the in-plane tensile properties of CXL cornea increased relatively less than the apparent connectivity between fibers.

To date, the specific relationship between corneal micro- and macro-structure with shape and focusing power remains speculative. Based on the NITI model, it can be hypothesized that a loss in lamellar inter-connectivity or degradation of the proteoglycan host-mesh would produce a higher degree of lamellar slippage while degradation in fiber strength along its axis would produce reduced tensile strength and a corresponding change in shape. In keratoconus progression, it has been shown that reduced interlamellar branching can be found in the anterior portion that corresponds to a measurable reduction in tensile modulus.[54] Reduced biomechanical stability due to lamellar slippage and collagen degradation has been suggested as a potential mechanism behind disease progression,[8,55–57] but remains difficult to monitor in a clinical setting. Still, studies of micro- and macro- biomechanical and biochemical stability of the cornea following CXL are required to better understand both short- and long-term effects on refractive changes.

Due to difficulties in acquiring human research tissue, all samples were scanned once the viable donor period had passed (greater than 1-2 weeks following enucleation). As such, biomechanical changes due to tissue necrosis and swelling may have occurred in samples prior to testing. Note that the apparent corneal stiffness (both $E$ and $G$) for different hydration levels has been explored in detail using numerous animal models.[33,42,46,58–60] In the present study, cross-linking via the mock Dresden protocol induced a change in thickness from approximately 577 μm to 450 μm, and a measured doubling of the Young's modulus. The doubling in tensile modulus was greater than what would be expected due to swelling alone. As reported previously, the shear modulus $G$ appeared less sensitive to hydration compared to the tensile modulus. In the samples tested in Group B, we measured a 3-fold increase in the shear modulus $G$; again, a change unlikely to result from thinning alone.

While carefully accounting for swelling in all samples remains an area of future study, the degree to which swelling alone, versus cross-linking alone, changes mechanical properties remains unclear. Of course, it remains likely that the reported differences in stiffness were due to a combination of simultaneous cross-linking and thinning. Because the present study sought to demonstrate that non-contact OCE can quantify changes induced in corneas undergoing a procedure similar to a clinical protocol, it does not measure the effect of cross-linking alone. Instead, a tool to measure expected changes during and following CXL is presented. Note that the measured differences in thickness generally agree with what has been reported elsewhere in both in vivo and ex vivo studies using a similar CXL procedure.[33,61–63]

Ideally, real-time monitoring of corneal stiffness during CXL would enable immediate feedback on procedure progression. In this study (Group C), we noted different rates of stiffening for individual corneas. We also note that there was a measurable difference in both moduli and thickness following the application of the riboflavin/dextran solution. Simply soaking the cornea in dextran can result in a measurable change in corneal thickness and stiffness,[33] consistent with the pre-CXL and pre-UV measurements in Group C. As corneas were exposed to UV illumination, OCE could measure the overall effect of the mock CXL procedure (from both Rf/Dextran soaking and UV-illumination) in the Dresden protocol. If the effects of both total stiffening and the rate of stiffening induced by Rf/Dextran and UV-illumination need to be explored independently, further studies are required.

Because research tissue was difficult to acquire, the inclusion of older aged samples (**Supplemental Methods 6**), may have resulted in generally higher moduli.[64] In fact, corneas from eye-banks tend to be from older donors, which makes it difficult to study samples from the age groups that most commonly present progressing keratoconus (i.e., age groups most likely to undergo CXL).[65] Still, due to a host of microstructural differences and unique cross-linking behaviors between human and animal models[36,37,66–69], even older human samples can provide valuable information and help demonstrate the potential for in vivo human use. Difficulties acquiring human tissues for research are not unique and have been recognized previously as a limiting factor in acquiring sufficient sample sizes.[65,67,69]

While the moduli quantified with AµT-OCE are generally consistent with both literature and mechanical testing, the proposed method has several limitations. One is that depth-dependent moduli cannot be measured using an approach based on guided mechanical waves and, consequently, the measurements most likely correspond to depth-averaged values of corneal moduli. It has been shown that CXL is most dramatic in the anterior portion of the cornea,[67] which contains a complex fiber arrangement that likely determines most of its flexural (tensile) strength. It has also been shown that the relative increase in $G$ is much greater in the anterior portion of the cornea.[46] Thus, if there was a significant increase in anterior cross-linking, much of the stroma may still be susceptible to inter-lamellar slippage. Exploring the importance of depth-resolved changes, as well as the interplay between moduli and intraocular pressure, remain directions of future interest.

Treatment success in CXL is currently defined by secondary measures such as corneal shape and refractive power changes over time.[70] To truly optimize outcomes, elastic moduli maps in cornea before and after surgery could evaluate individual stiffness correction requirements and monitor whether the required correction had been reached. As focal biomechanical weakening in the cone can occur in progressive keratoconus, localized biomechanical mapping is more likely to provide the information needed to develop a personalized model for each individual cornea suitable for screening, surgical planning, and treatment monitoring. Because elastic moduli maps may be combined with topography measurements provided by ophthalmic systems, topographic and mechanical changes would not only provide potentially important diagnostic information but may also be used to monitor localized cross-linking treatments. Although OCT can provide micron-scale resolution in tracking mechanical waves, the resultant OCE resolution in mapping mechanical properties across the cornea is lower.[28,29] Cornea is a bounded material and, therefore, guided mechanical waves further degrade the resolution in mapping moduli. While high resolution images of local group velocity have been demonstrated in 3-dimensional corneal sections,[26] local mapping of elastic moduli has not yet been demonstrated. Methods improving spatial resolution in dynamic OCE measurements of bounded material (such as the cornea) are currently under investigation, and the demonstration of spatially resolved maps of cornea mechanical moduli is a subject of future studies.

The standard Dresden protocol involves a 30-min corneal exposure to 370 nm UV radiation at an irradiance of 3 mW/cm$^2$ following riboflavin/dextran saturation.[45] Several other protocols have been proposed which reduce exposure time. For example, accelerated CXL uses higher irradiance UV-A to shorten the time needed to deliver the equivalent total energy dose.[17] While different protocols have demonstrated varying degrees of stiffening based on the time of UV irradiance,[40] their translation remains limited and cannot be personalized without quantitative measurements of corneal elasticity, which is critical in guiding CXL treatment and predicting outcomes.

Finally, we note that thickness remains an important parameter in determining whether or not to perform CXL based on the hypothesis that riboflavin serves as both a cross-linking agent and a means to protect the posterior segment of the eye from UV radiation.[71–78] While analytical models describing riboflavin loading, UV light exposure, and polymerization have been used to determine safety limits and predict surgical outcomes, current treatment parameters are designed assuming that a large majority of the UV light is absorbed or scattered in the anterior 250 to 350 μm of the corneal stroma in all patients. Thus, a minimum stromal thickness of 400 μm is required to protect corneal endothelium and deeper ocular structures.[76] This approach often excludes advanced keratoconus patients. We hypothesize that OCE can be used to design patient-specific procedures balancing the benefit between UV-radiation time and desired biomechanical stabilization, opening the possibility for optimized CXL treatment of even advanced keratoconus patients.

In this study, we demonstrate the potential of AμT-OCE to monitor stiffening and thinning of the cornea intraoperatively, with varying rates of change in $E$ and $G$. Although we demonstrated the potential to quantify both elastic moduli non-invasively using a non-contact near real-time procedure, it is important to note that results obtained in ex vivo studies (such as the one reported herein) may not fully correspond to changes expected in in vivo corneas using the same protocol. Given the favorable ectasia-stabilizing results with cornea cross-linking therapy, efforts aimed at optimizing the procedure's treatment time, intra- and postoperative comfort, and efficacy remain an area of focus.

**Conclusions**

Corneal cross-linking is increasingly being used clinically to biomechanically stabilize the cornea in vision degrading diseases such as keratoconus or post-LASIK ectasia. While clinical outcomes are largely positive, there remains a gap in both understanding the mechanisms behind corneal stabilization and providing clinicians feedback on how those mechanisms may relate to disease progression and treatment plans.

In this study, AµT-OCE quantified changes in both the in-plane Young's modulus, $E$, and out-of-plane shear modulus, $G$, of human donor cornea due to UV induced crosslinking in a non-contact, non-destructive manner. The results demonstrated a significantly different stiffening response in both $E$ and $G$, where $G$ experienced a larger relative increase from the cross-linking procedure. This suggests that while CXL strengthened the tensile and flexural strength of the tissue, the ability of the cornea to resist shearing forces was more dramatically altered by the procedure.

AµT-OCE provides a unique non-invasive and non-contact tool that can monitor and evaluate corneal elasticity, including changes induced by UV-CXL. CXL-induced changes in mechanical anisotropy likely carry diagnostic and prognostic information that warrant future studies.

**References**

1. Meek, K. M. & Boote, C. The organization of collagen in the corneal stroma. *Exp. Eye Res.* **78**, 503–512 (2004).

2. Koudouna, E. *et al.* Evolution of the vertebrate corneal stroma. *Prog. Retin. Eye Res.* **64**, 65–76 (2018).

3. Borcherding, M. S. *et al.* Proteoglycans and collagen fibre organization in human corneoscleral tissue. *Exp. Eye Res.* **21**, 59–70 (1975).

4. Scott, J. E. & Thomlinson, A. M. The structure of interfibrillar proteoglycan bridges ('shape modules') in extracellular matrix of fibrous connective tissues and their stability in various chemical environments. *J. Anat.* **192**, 391–405 (1998).

5. Meek, K. M. & Knupp, C. Corneal structure and transparency. *Prog. Retin. Eye Res.* **49**, 1–16 (2015).

6. Gandhi, S. & Jain, S. The anatomy and physiology of cornea. *Keratoprostheses Artif. Corneas Fundam. Surg. Appl.* **37**, 19–25 (2015).

7. Kotecha, A. What Biomechanical Properties of the Cornea Are Relevant for the Clinician ? *Surv. Ophthalmol.* **52**, 109–114 (2007).

8. Dupps, W. J. & Wilson, S. E. Biomechanics and wound healing in the cornea. *Exp. Eye Res.* **83**, 709–720 (2006).

9. Andreassen, T. T., Hjorth Simonsen, A. & Oxlund, H. Biomechanical Properties of Keratoconus and Normal Corneas. *Exp. Eye Res* **31**, 435–441 (1980).

10. Chan, E. & Snibson, G. R. Current status of corneal collagen cross-linking for keratoconus: A review. *Clin. Exp. Optom.* **96**, 155–164 (2013).

11. Santhiago, M. R. & Randleman, J. B. The biology of corneal cross-linking derived from ultraviolet

Supplemental Online Content

Non-contact acoustic micro-tapping optical coherence elastography for evaluating biomechanical changes in the cornea following UV/riboflavin collagen cross linking: ex vivo human study

Mitchell A. Kirby[1], Ivan Pelivanov[1], Gabriel Regnault[1], John J. Pitre[1], Ryan T. Wallace[2], Matthew O'Donnell[1], Ruikang K. Wang[1,3], Tueng T. Shen[3, *]

[1]Department of Bioengineering, University of Washington, Seattle, Washington 98105, USA
[2]School of Medicine, University of Washington, Seattle, Washington 98195, USA
[3]Department of Ophthalmology, University of Washington, Seattle, Washington 98104, USA
*Correspondence: Tueng T. Shen, UW Medicine Eye Institute, Seattle, Washington 98104, USA.
Email: ttshen@uw.edu

**Supplementary Methods**

**1. Estimating goodness of fit**

An iterative routine to estimate in-plane tensile and out-of-plane shear moduli ($\mu$ and $G$, respectively) was previously presented.[1] The method finds the theoretical $A_0$ dispersion mode of a guided elastic wave in a nearly-incompressible transverse isotropic (NITI) material that most closely matches experimentally measured elastic wave propagation.[1,2] Iteration was performed in the frequency-wavenumber (*f-k*) domain using simplex optimization (*fminsearch*, MATLAB, MathWorks, Natick, MA), where the best-fit theoretical dispersion relation was performed by maximizing the following objective function:

$$\Phi(\mu, G) = \frac{1}{N_f} \sum_f \sum_k w(f, k; \mu, G) |\hat{v}(f, k)|^2 - \beta \left|\frac{\mu}{\lambda}\right| \quad (S.1)$$

where $\hat{v}$ is the normalized 2D Fourier spectrum of the measured time-space distribution of vertical displacements induced by the propagating guided wave. The function $w(f, k; \mu, G)$ is related to the $A_0$ mode solution for a NITI ($G < \mu$) or isotropic ($G = \mu$) material. The regularization term,

$\beta$, satisfying the nearly-incompressible assumption, was set to 1, based on an L-curve analysis.[2] The value of $\lambda$ was updated at each iteration according to $\lambda = \rho c_L^2 - 2\mu$. Note that prior to computing the 2D Fourier spectrum, a temporal super Gaussian filter ($SG$) that followed the maximum vibration velocity of the wave-field $t_m^{wf}(x)$ at each discrete position $x$, was applied:

$$SG(t) = \exp\left[-\left(\frac{1}{2}\left(\frac{t - t_m^{wf}(x)}{\sigma_t}\right)^2\right)^2\right] \tag{S.2}$$

where $\sigma_t = 0.5$ ms. This step was included to prevent error in the fitting routine due to low frequency noise induced by the experimental setup.

The optimization function was not posed as a least-squares error problem. Rather, the procedure sought to maximize the spectral energy contained in a small, weighted window around the $A_0$ mode. At a given frequency $f$ and current parameter set $(\mu, G)$, the dispersion relation solver returned the wavenumber associated with the $A_0$ mode, $k_0$. The weighted window function at each frequency centered at the corresponding $k_0$ was defined by $w(f, k; \mu, G)$, where $\Delta k$ was the discrete wavenumber sampling interval:

$$w(f, k; \mu, G) = \begin{cases} e^{-\frac{1}{2}\left(\frac{k - k_0(f, \mu, G)}{1.2\Delta k}\right)^2}, & |k - k_0(f, \mu, G)| \leq 3\Delta k \\ 0, & \text{otherwise} \end{cases} \tag{S.3}$$

where the weights were assigned at each frequency with a peak value at the wavenumber of the theoretical dispersion curve and weights decaying (with a Gaussian distribution) with distance from the theoretical curve.

The dispersion relation estimated the location of the $A_0$ mode (e.g. the wavenumber of the mode for a given frequency) but not the spectral energy in a frequency-wavenumber bin (which depends on the method used to excite guided waves). Essentially, this approach found the dispersion curve that overlapped with the maximum spectral power. Multiple physical parameters were considered fixed, including the corneal density ($\rho$ = 1000 kg/m³), corneal longitudinal wave speed ($c_L$ = 1540 m/s), and mean corneal thickness (measured from B-mode OCT images). The cornea was assumed bounded from below by water with a density of 1000 kg/m³ and longitudinal wave speed of 1540 m/s. To avoid convergence to a local (as opposed to global) maximum in equation S.1, five independent fits were performed, with quasi-random initial values of $G_0$ and $\mu_0$. The final output of $\mu$ and $G$ were set to those corresponding to the highest value in equation S.1.

To quantify how well the NITI solution fits experimental data, a goodness-of-fit (GOF) metric was introduced based on an "unconstrained, global optimum" for the objective function, $\Phi_{max}$. At each frequency $f$, the following optimization problem was solved:

$$k_{max} = \underset{\tilde{k}}{\mathrm{argmax}} \sum_{k} \tilde{w}(f,k;\tilde{k},\mu,G)|\hat{v}(f,k)|^2 \qquad (S.4)$$

$$\tilde{w}(f,k;\tilde{k},\mu,G) = \begin{cases} e^{-\frac{1}{2}\left(\frac{k-\tilde{k}}{1.2\Delta k}\right)^2}, & |k-\tilde{k}| \leq 3\Delta k \\ 0, & \text{otherwise} \end{cases} \qquad (S.5)$$

$$\Phi_{max}(\mu,G) = \frac{1}{N_f} \sum_{f} \sum_{k} \tilde{w}(f,k;k_{max},\mu,G)|\hat{v}(f,k)|^2 \qquad (S.6)$$

This was equivalent to finding the Gaussian-weighted window containing the most spectral energy at each frequency independently and summing those contributions to the total mode energy. Since

$k_{max}$ may vary at each frequency independent of the dispersion relation, $k_{max}$ may not follow the mode shape exactly and may not even be smooth. However, this means that $\Phi_{max}$ represents an upper bound on the value of $\Phi$ calculated during fitting. We therefore produced the following GOF metric:

$$g_{NITI} = \frac{\Phi_{NITI}}{\Phi_{max}} \tag{S.7}$$

$\Phi_{NITI}$ covered the maximum energy with unique combinations of $\mu$ and $G$, constrained by the shape of the $A_0$ dispersion curve in the NITI model, while $\Phi_{max}$ captured the unconstrained energy. Thus, $g_{NITI}$ indicated the portion of the maximum possible mode energy captured by a given $A_0$ dispersion curve. Values near $g_{NITI} <\approx 1$ suggest that the theoretical dispersion curve is well described by the NITI model, where $\Phi_{NITI}$ captures nearly all the measured mode's energy.

## 2. Quantifying uncertainty intervals

While $g_{NITI}$ provides an estimate for how well the theoretical $A_0$ mode matches experimental data, it alone does not provide confidence intervals on the output moduli $G$ and $\mu$. Due to the maximization approach for weighted fitting based on the energy in the 2-D Fourier transform, residual errors are not computed, making traditional confidence interval methods difficult to apply. For example, a low value of $g_{NITI}$ would suggest that the $A_0$ dispersion curve calculated from the NITI model poorly described the actual dispersion measured within the sample. In such case, modulus estimates should have increased uncertainty.

To estimate uncertainty, $G$ and $\mu$ were varied independently around the values near a maximum $g_{NITI}$ and $\Phi(\mu, G)$ was recorded. For each combination of $G$ and $\mu$, Eq. S.1 was used to calculate $\Phi(\mu, G)$, which was then used to determine:

$$\psi(\mu, G) = \frac{\Phi(\mu, G)}{\Phi_{max}}, \tag{S.8}$$

where the $\psi$ function represented goodness of fit values normalized to the maximum energy for a range of values. A representative example of $\psi(\mu, G)$ can be seen in **Fig. S1.** Here (XT-plot shown in **Fig. S1a**), the iterative routine (S. 1) converged on a best-fit $A_0$ mode (**Fig. S1b**) where $g_{NITI} = 0.98$ when $G = 70$ kPa and $\mu = 1.5$ MPa. In **Fig. S1a**, $\psi$ is shown when $\mu = 1.5$ MPa and $G$ was varied from 40 kPa- 130 kPa. The corresponding $A_0$ dispersion curves (constitutive equation found in Ref[2]) as $G$ varied can be seen in **Fig. S1b**. As described previously, the high-frequency threshold of the $A_0$ dispersion curve is largely determined by $G$. In **Fig. S1c**, $\psi$ is shown when $G = 70$ kPa and $\mu$ swept across a 0.1 MPa - 15 MPa range. Again, the corresponding $A_0$ dispersion curves can be seen in **Fig. S1d**. As suggested previously, the $A_0$ mode is not as sensitive to changes in $\mu$ for the degree of shear anisotropy ($\mu/G$) expected in the cornea. As such, $\psi$ is less sensitive to changes in $\mu$ and produces higher relative uncertainty. This routine provided a range for both $G$ and $\mu$ values indicating the degree to which the iterative solution converged on a single value.

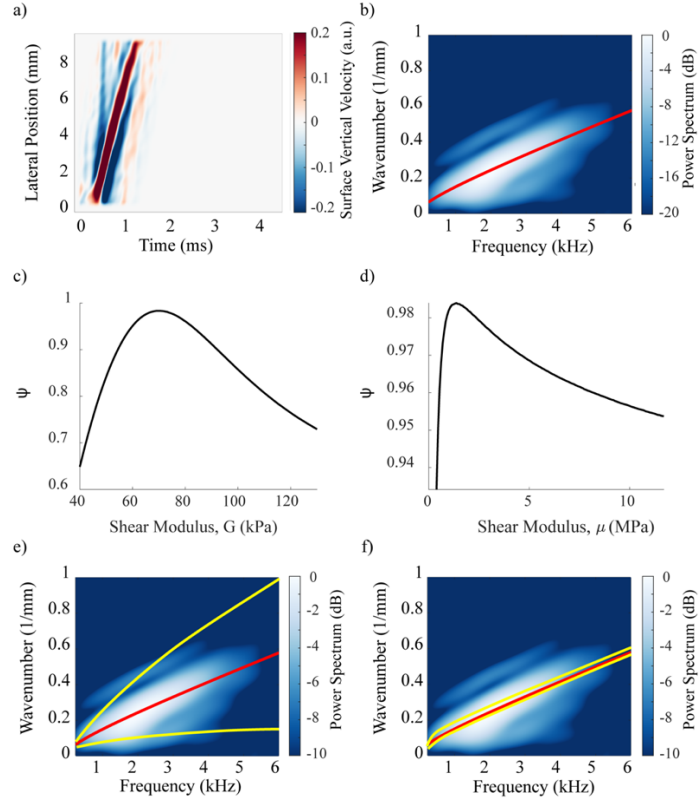

**Figure S1.** a) XT-plot of surface vibrations measured via OCT in a normal cornea sample. b) Best-fit solution to the dispersion equation (based on a unique combination of elastic moduli, $\mu$ and $G$, displayed in red) on top of the measured waveform in the 2-D Fourier spectrum for the corresponding cornea c) $\psi$ for $\mu = 1.5$ MPa, where $G$ was swept across a range from 40 kPa- 130 kPa. d) $\psi$ for $G = 70$ kPa, where $\mu$ was swept across a range from 0.1 MPa- 15 MPa. Yellow lines plotted on top of measured energy in the ($f$-$k$) domain are $A_0$ dispersion curves corresponding to the range e) $G = 40$ kPa and $G = 150$ kPa, with $\mu = 1.5$ MPa and f) $\mu = 0.1$ MPa and $\mu = 15$ MPa, with $G = 70$ kPa.

To determine the range of uncertainty, an approach estimating both model and data error was developed. To account for data (noise) error, n=5 repeat scans were taken in each cornea at each IOP, and $g_{\text{NITI}}$ was determined independently for each scan. In the representative example shown above, the mean value of $g_{NITI}$ ($g_{NITI} = \psi_{max}$) was 0.98 with a standard error of 0.02 (or 2%) across 5 independent scans. The variation in $g_{NITI}$ determined the uncertainty interval using the corresponding $G$ and $\mu$ values that resulted in $\psi$ being 2% lower than the max (i.e. $\psi = .96$ for $G$ and $\mu$) (**Fig. S2**). Due to the shape of $\psi$, the uncertainty in the fit produced uneven error bars. Note that the absolute value of $g_{NITI}$ provides an estimate for model error, where $g_{NITI} = 1$ for a NITI material would have very small uncertainty intervals. As $g_{NITI}$ is reduced, the shape of $\psi$ widens for both moduli and model uncertainty increases.

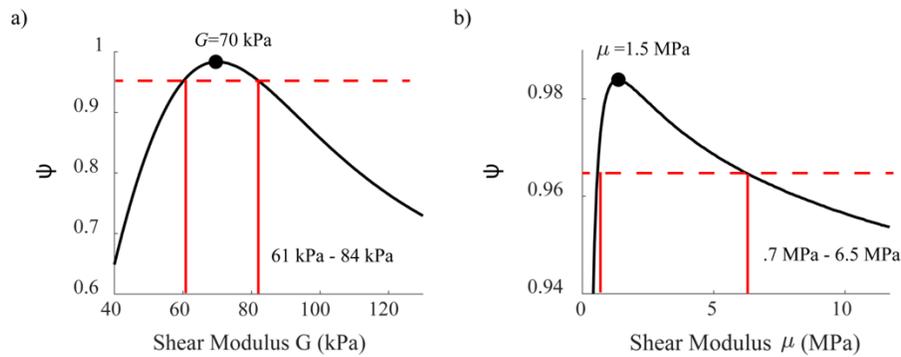

**Figure S2.** Uncertainty intervals (red line) calculated for the representative example where $g_{NITI}$ standard error was 2%. The black dotted line is $\psi$ determined by (S. 8). a) The best fit solution for this scan was $G = 70$ kPa, with uncertainty of 61 kPa – 84 kPa. b) The best fit solution for this scan was $\mu = 1.5$ MPa, with uncertainty of 0.7 MPa – 6.5 MPa.

Because n=5 repeat scans were taken, five independent measurements produced corresponding values for $G$ and $\mu$ and their respective uncertainty ranges. The uncertainty ranges for each cornea (at each IOP) were finally calculated by taking the square-root of the means of the upper and lower limits, divided by the number of scans (n=5). Note that these intervals are displayed in **Figure S6, Figure S7,** and **Figure S10,** as well as in **Figure 7** in the main paper.

## 3. Exclusion criteria

In some cases, the NITI model does not describe measured wavefields properly (due to poor excitation, misalignment, corneal structure abnormalities, etc.). To determine which scans can be described by the NITI model, a 'cut-off' criterion in the goodness of fit for both $G$ and $\mu$ was determined. Any scan with a goodness of fit below the cut-off value was omitted from analysis.

To determine the relationship between $g_{\text{NITI}}$ and model error in human corneas, an iterative fit was performed and $g_{\text{NITI}}$ calculated for untreated (**Figs. S3a-b**) and crosslinked (**Figs. S4a-b**) corneas separately, providing $g_{\text{NITI}}$/moduli histograms. Data from both group A and B are used here. Each section of the histograms corresponds to a 0.01 range in $g_{\text{NITI}}$ and, respectively, 20 kPa for $G$ and 2 MPa for $\mu$. These histograms illustrate the repartition of moduli obtained from the fits and show, as expected for a given group, a stable repartition of moduli in the high goodness of fit range (for a given group the stiffness moduli should be easily within an order of magnitude between every individual).

In order to determine the critical $g_{\text{NITI}}$ value, the histograms were integrated in the $g_{\text{NITI}}$ direction. For every range of stiffness modulus, the number of occurrences was counted and the mean $g_{\text{NITI}}$ computed. These results are shown in **Figs. S3c-d** and **S4c-d** for, respectively, untreated and crosslinked cases. The modulus clearly increases (associated with faster wave speeds) as $g_{\text{NITI}}$ decreases progressively before approaching a point where this behavior breaks, reaching a plateau. The plateau is associated with nearly random statistics for reconstructed moduli, corresponding to the range of $g_{\text{NITI}}$ below which the fitting procedure used for moduli reconstruction is inaccurate. The final cutoff value was computed as the average $g_{\text{NITI}}$ over the plateau range. **Table S1** presents the cutoff goodness of fit ($g_{NITI}$) criteria determined as explained above for both untreated and CXL corneas, and for both $G$ and $\mu$.

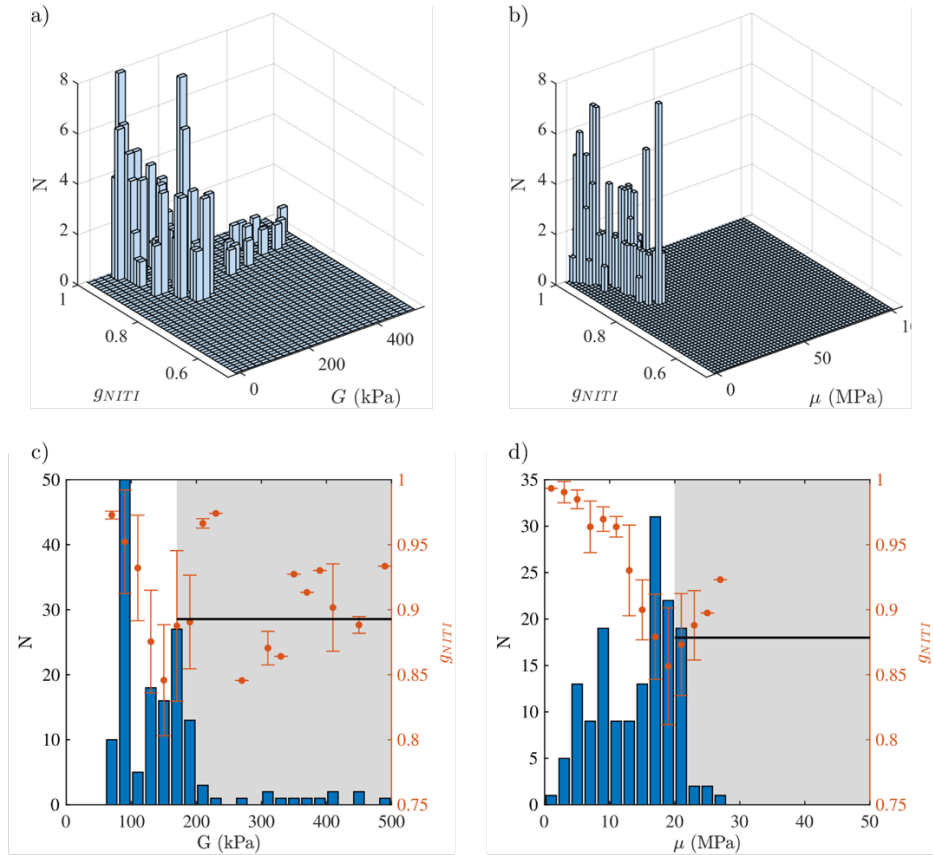

**Figure S3.** Procedure to determine the criteria of exclusion for $G$ and $\mu$ for untreated corneas. a), b) 2D histograms illustrating the distribution of fitted stiffness moduli, respectively for $G$ and $\mu$, and the goodness of fit, $g_{NITI}$, metric. c), d) 1D histogram illustrating the distribution of data as a function of the fitted stiffness moduli, respectively for $G$ and $\mu$. The right axis represents the averaged $g_{NITI}$ in the considered modulus range and the error bars represent the standard deviation in the measured goodness of fit. The black line and gray shaded area indicate the range over which the cut off $g_{NITI}$ was calculated. The exact values for the cut off $g_{NITI}$ are given in **Table S1**).

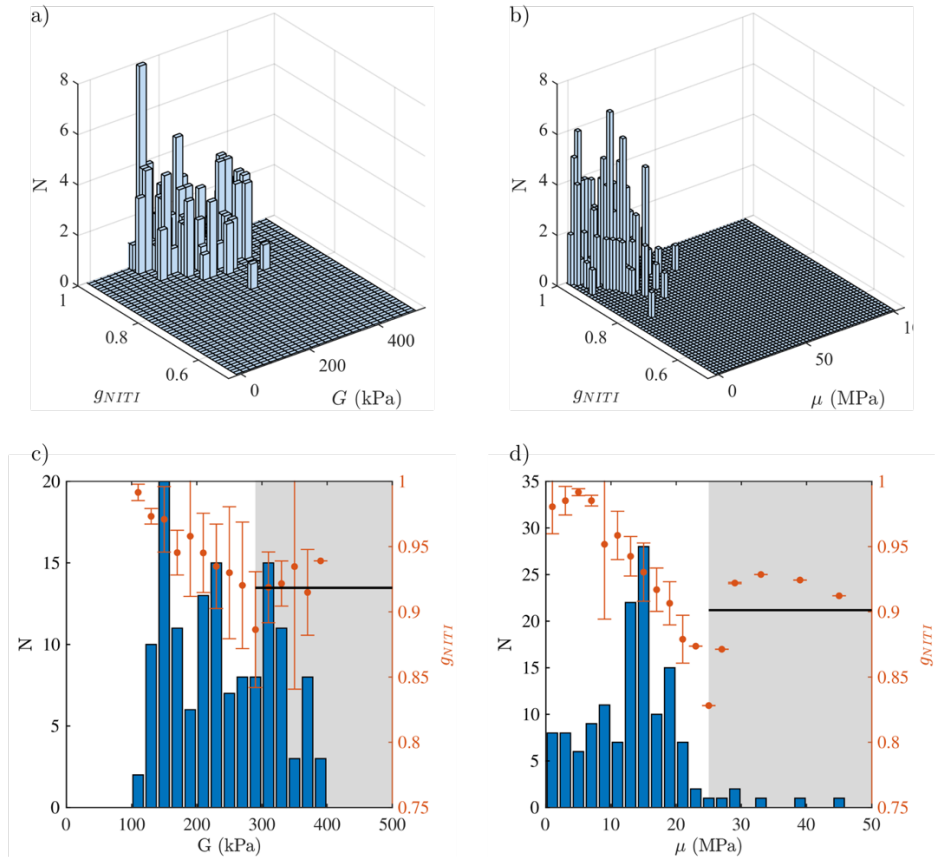

**Figure S4.** Procedure to determine the exclusion criteria for $G$ and $\mu$ in crosslinked corneas. a), b) 2D histograms illustrating the distribution of fitted moduli, respectively for $G$ and $\mu$, and the goodness of fit, $g_{NITI}$, metric. c), d) 1D histogram illustrating the distribution of data as a function of the fitted moduli, respectively for $G$ and $\mu$. The right axis represents the averaged $g_{NITI}$ in the considered modulus range and the error bars represent the standard deviation in the measured goodness of fit. The black line and gray shade indicate the range over which the cut off $g_{NITI}$ was calculated. The exact values for the cut off $g_{NITI}$ are given in **Table S1**).

**Table S1:** Cut off values of goodness of fit, $g_{NITI}$, for untreated and crosslinked corneas

| Modulus | Untreated corneas | Crosslinked corneas |
|---------|-------------------|---------------------|
| $G$     | 0.89              | 0.9                 |
| $\mu$   | 0.88              | 0.9                 |

In Group A, 9 of the 100 total scans produced output quality below the exclusion criteria for $G$ in the untreated (A1) group, where 6 of the 9 poor quality fits were at 20 mmHg, 1 at 15 mmHg, and 2 at 10 mmHg. For $\mu$ in the untreated group (A1), 11 of the 100 total scans provided fit quality below the exclusion criteria, where 7 of the 11 poor quality fits were at 20 mmHg, 2 at 15 mmHg, and 2 at 10 mmHg.

In the CXL group (A2), 18 of the 100 total scans had fit quality beyond the exclusion criteria for $G$, where 5 exclusions occurred in corneas inflated to 20 mmHg, 7 at 15 mmHg, 4 at 10mmHg, and 2 at 5 mmHg. For $\mu$ in the CXL group, 34 of the 100 total scans provided fit quality below the exclusion criteria where 10 of the 34 exclusions occurred at 20 mmHg, 11 at 15 mmHg, 5 at 10 mmHg, and 8 at 5 mmHg.

For Group B, in the full dataset of 70 scans, 14 scans were omitted from the analysis due to poor fit quality. Twelve (12) of the 14 scans with poor fit quality were in the samples following CXL. For Group C, approximately 20% of the scans (30 out of 160) corresponded to poor fit quality.

### 4. Cornea data

In this work, human samples were probed via AµT-OCE to simultaneously quantify both elastic moduli ($G$ and $E$, where $E = 3\mu$) in artificially inflated cornea. Both $g_{NITI}$ and $\psi$ were determined in each scan, exclusion criteria were applied, and uncertainty intervals associated with each moduli determined.

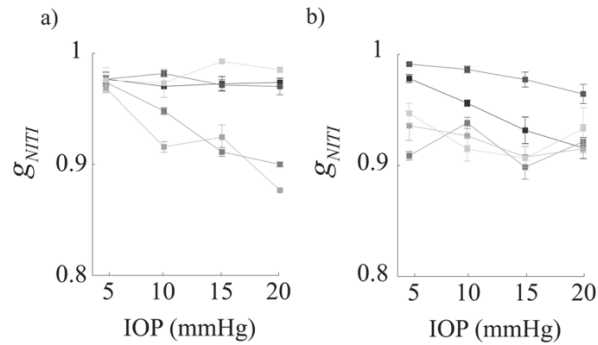

**Figure S5.** Goodness of fit ($g_{NITI}$) for cornea in the a) untreated group (A1) and b) CXL group (A2). Error bars correspond with the standard deviation among five repeat scans on each cornea. Different shades represent different independent samples.

Within Group A, two independent groups ('untreated' group (A1) and 'CXL' group (A2)) were scanned with AμT-OCE. The values for $g_{NITI}$ in both sub-groups can be seen in **Fig. S5**. The values for $G$ and $E$ from both OCE and mechanical testing are shown in the untreated group (A1) (**Fig. S6**) and cross-linked group (A2) (**Fig. S7**). In Group A, note that as the value of $g_{NITI}$ decreased, uncertainty in the modulus generally increased. In the untreated group, all scans for one sample at 20 mmHg were below the omission criteria.

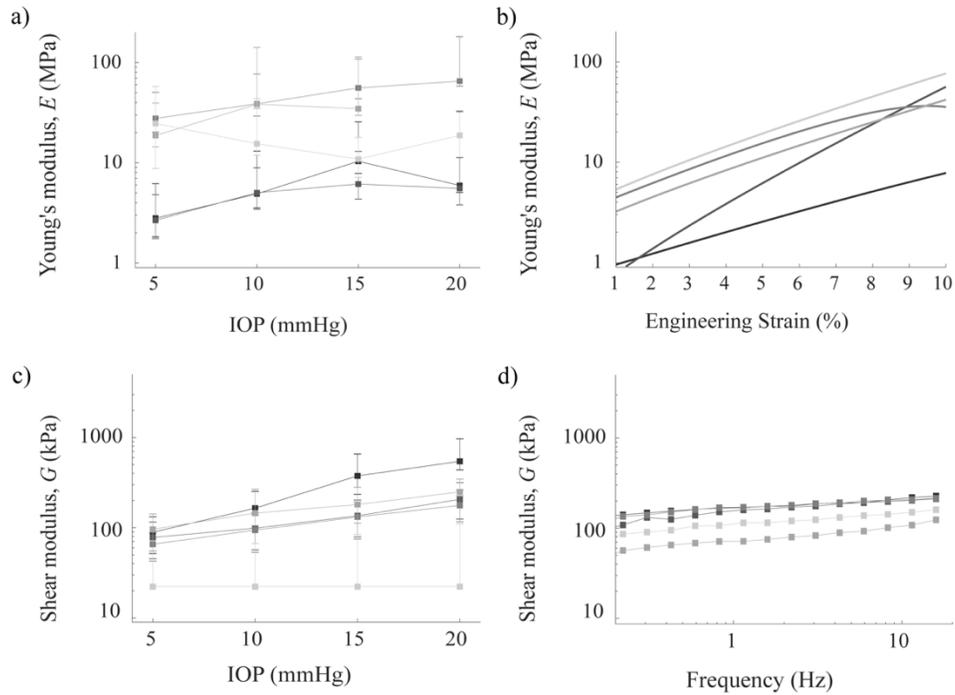

**Figure S6.** Mechanical test results in untreated (A1) cornea group. a) In-plane Young's modulus $E$ measured with AμT-OCE at intraocular pressures from 5-20 mmHg. b) Strain-dependent Young's moduli, $E$, measured via extension testing up to 10% strain, or where visible tissue damage occurred. c) Out-of-plane shear modulus, $G$, measured with AμT-OCE at intraocular pressures from 5-20 mmHg. d) Out-of-plane shear modulus, $G$, measured with parallel plate rheometry. Each shade of gray represents the same cornea.

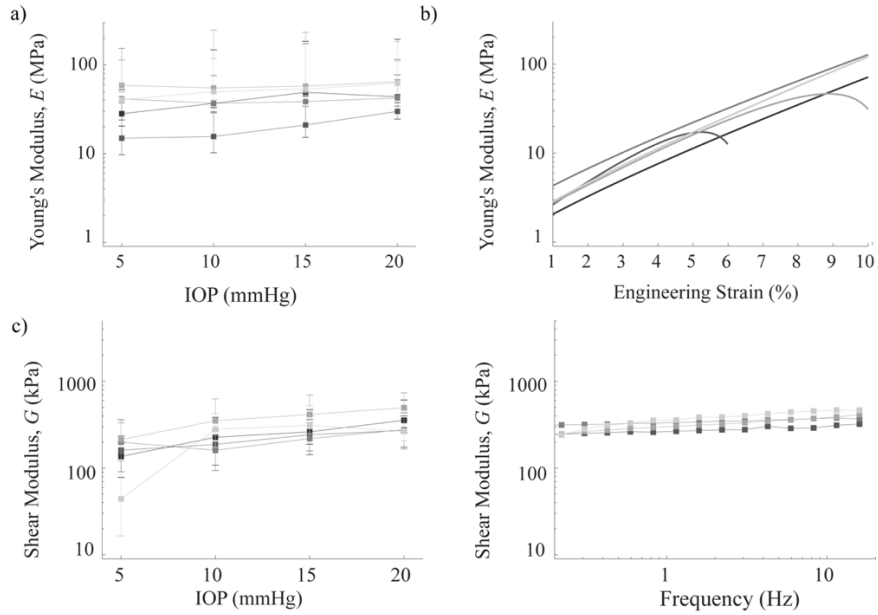

**Figure S7.** Mechanical test results in CXL (A2) cornea group. a) In-plane Young's modulus, $E$, measured with AμT-OCE at intraocular pressures from 5-20 mmHg. b) Strain-dependent Young's moduli, $E$, measured via extension testing up to 10% strain, or where visible tissue damage occurred. c) Out-of-plane shear modulus, $G$, measured with AμT-OCE at intraocular pressures from 5-20 mmHg. d) Out-of-plane shear modulus, $G$, measured with parallel plate rheometry. Each shade of gray represents the same cornea.

In group B, moduli pre- and post- CXL were measured via OCE and reported in the results section of the main paper. The corresponding goodness of fit values are displayed in **Fig. S8**.

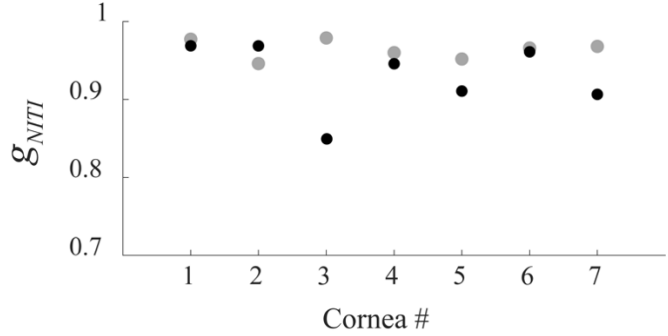

**Figure S8.** Goodness of fit ($g_{NITI}$) before (gray) and after (black) CXL. Note that for all but one sample, $g_{NITI}$ decreased following CXL. Each shade of gray represents the same cornea.

For the samples in Group C (see **Methods**) the intra-procedure scan (i.e. OCE during CXL) was only performed once at each time point, as opposed to the 5 repeat scans described above. The goodness of fit, $g_{NITI}$, for each scan is shown in **Fig. S9**.

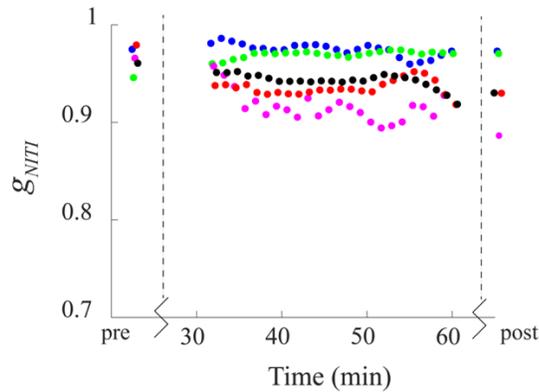

**Figure S9** a) Goodness of fit, $g_{NITI}$, values for each cornea, before, at time points during the mock CXL procedure, and after. Generally, fit quality decreased over time.

To determine uncertainty intervals, the standard error in $g_{NITI}$ across the entire procedure was used to determine the cut-off in $\psi$ for each scan. (**Fig. S10**).

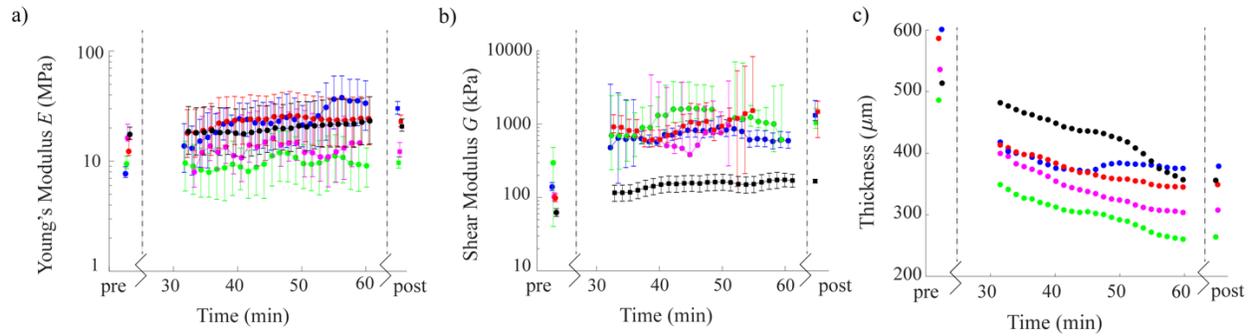

**Figure S10:** a) OCE determined $E$ and b) $G$ before, during, and following CXL. Error bars represent individual uncertainty intervals. c) corneal thickness before, during, and following CXL. Each color corresponds to the same sample.

Relative to untreated corneas, samples that underwent CXL generally had a lower $g_{NITI}$, and, as such, larger uncertainty. The increased uncertainty is likely due to both increased noise and model error. As CXL increases stiffness, there will inherently be a trade-off with AµT-generated elastic wave amplitude. In this case, the signal-to-noise ratio (SNR) will be reduced. In addition to SNR-induced error, collagen cross-linking may physically alter the lamellar structure in a way that renders the NITI model less accurate. We speculate that the CXL procedure may also lead to depth-dependent stiffening of the corneal structure due to a limited penetration of both riboflavin solution and UV radiation inside the cornea. As such, the derivation of the uncertainty presented here takes both sources of error into account and estimates uncertainty accordingly.

There can be an additional reason for NITI model inaccuracy. First, as we reported previously, at IOP higher than 15 mmHg, the NITI model can start to fail due to geometrical deformation of the eyeball resulting in an additional lateral anisotropy, which makes an orthotropic model more appropriate to describe corneal elasticity.[2,3]

## 6. Donor information

| Tissue Information | | | | |
|---|---|---|---|---|
| Group | Age | Gender | Race | Time from Preservation (days) |
| Group A Normal | | | | |
| | 48 | F | Caucasian | 40 |
| | 55 | M | Caucasian | 44 |
| | 61 | M | Caucasian | 20 |
| | 55 | - | - | 48 |
| | 48 | M | Caucasian | 7 |
| Group A Cross-Linked | | | | |
| | 23 | - | - | 21 |
| | 23 | - | - | 21 |
| | 46 | M | Caucasian | 10 |
| | 26 | M | Caucasian | 12 |
| | 26 | M | Caucasian | 12 |
| Group B | | | | |
| | 21 | M | Caucasian | 29 |
| | 45 | F | Caucasian | 35 |
| | 62 | M | Caucasian | 25 |
| | 62 | M | Caucasian | 25 |
| | 48 | M | Caucasian | 28 |
| | 49 | F | Caucasian | 37 |
| | 73 | F | Asian | 33 |
| Group C | | | | |
| | 66 | M | Caucasian | 16 |
| | 66 | M | Caucasian | 16 |
| | 59 | F | Caucasian | 40 |
| | 68 | M | Caucasian | 50 |
| | 49 | F | Caucasian | 35 |